\documentclass{appolb}
\usepackage{epsfig}
\usepackage{graphicx}
\usepackage{dcolumn}   % needed for some tables
\usepackage{bm}        % for math
\usepackage{amsmath}
\usepackage{amssymb}   % for math
\usepackage{enumitem}
\usepackage{graphicx,mathrsfs}
\usepackage{nicefrac}
\usepackage{multirow}
\usepackage{color}

\def\ba{\begin{eqnarray}}
\def\ea{\end{eqnarray}}

\begin{document}
% \eqsec  % uncomment this line to get equations numbered by (sec.num)
\title{A covariant model for the nucleon spin structure
\thanks{Presented at the Summer School and Workshop on High Energy Physics at 
the LHC: New trends in HEP, October 21- November 6 2014, 
Natal, Brazil}
}
\author{
G. Ramalho
\address{International Institute of Physics, Federal 
University of Rio Grande do Norte, Av.~Odilon Gomes de Lima 1722, 
Capim Macio, Natal-RN 59078-400, Brazil}
}
\maketitle
\begin{abstract}
We present the results of the covariant spectator quark model 
applied to the nucleon structure function $f(x)$ measured 
in unpolarized deep inelastic scattering, 
and the structure functions $g_1(x)$ and $g_2(x)$
measured in deep inelastic scattering using polarized beams
and targets  
($x$ is the Bjorken scaling variable).
The nucleon is modeled by a valence quark-diquark structure
with $S,P$ and $D$ components.
The shape of the wave functions and the 
relative strength of each component are 
fixed by making fits to the deep inelastic scattering data 
for the structure functions $f(x)$ and  $g_1(x)$.
The model is then used to make predictions on the function
$g_2(x)$ for the proton and neutron.
\end{abstract}

%Here comes the test, tables, figures....

%\hspace{.2cm}
\hspace{.1cm}

The covariant spectator quark model (CSQM) 
is a model in which the electromagnetic 
structure of the constituent quark
is parametrized by Dirac ($f_{1q}$) and 
Pauli ($f_{2q}$)  form factors for the quarks ($q=u,d$) 
\cite{Nucleon,Omega}.
The quark electromagnetic form factors $f_{1q}, f_{2q}$
simulate the effects associated 
with the gluons and the quark-antiquark pairs.
The CSQM was developed within 
the covariant spectator theory~\cite{Gross69}
and was first applied to the nucleon 
using a $S$-state approximation to the quark-diquark system~\cite{Nucleon}.
The quark form factors and the radial wave functions 
are fitted to the nucleon electromagnetic form factor data.
It was concluded that the 
falloff of the ratio between the magnetic and electric 
observed for the first time 
at  Jefferson Lab 
can be explained by 
a model based on quarks with 
no orbital momentum, if the quarks have 
an internal structure~\cite{Nucleon}.
%It was concluded that the novel results 
%from the Jefferson Lab {\bf showing that} the 
%observed falloff of the ratio between the magnetic and electric 
%form factors of the proton can be explained by 
%a model based on quarks with 
%no orbital momentum, if the quarks have 
%an internal structure~\cite{Nucleon}.
%
The model was later extended to several nucleon resonances
and other baryons~\cite{Omega,NSTAR,SQTM,Delta,Resonances,Octet}.

The next step on this, it is to check if CSQM 
can be extended to the  deep inelastic scattering (DIS) regime,
and if a qualitative description of the DIS 
phenomenology can be achieved.
In the deep inelastic scattering the photon 
transfer momentum squared, $Q^2$, and the photon energy in the lab frame, 
$\nu$, are both very large but the ratio $x=\frac{Q^2}{2M \nu}$
is kept finite  ($M$ is the nucleon mass).  
If the CSQM is in fact compatible with DIS,
the DIS data can be used to discriminate the 
individual contributions of the 
orbital angular momentum states 
in the nucleon wave function and %further
also  used to estimate 
the shape of those components.

The nucleon structure in DIS 
is parametrized in terms 
of the unpolarized structure functions $f_q(x)= q(x)$
and the polarized structure functions 
$g_1^q(x) = \Delta q (x)$ and $g_2^q(x)$.
The unpolarized structure functions 
determine the quark contributions to the nucleon momentum, but explain 
only about 50\% of the total amount
($\int dx (2 xf_u + xf_d) \approx 0.5$).
The remaining  50\% are due to the  gluons.
The functions $\Delta q$ measures the 
contributions to the quark orbital momentum 
for the proton spin.
It is known since the 80s, from the EMC experiments 
at CERN \cite{EMC},
that the contribution of the orbital momentum 
of the quarks to the  proton spin is only about 30\%~\cite{EMC,DIS-data}.
That conclusion was obtained from the result 
of the first moment of the function $g_1(x)$ 
for the proton~\cite{Kuhn09,DIS} 
\ba
\Gamma_1^p= \int_{0}^1 dx \, 
g_{1p}^{\rm exp}(x) = 0.128 \pm 0.013.
\ea
Theoretical calculations based on 
the naive assumption that the nucleon 
is made of quarks with no orbital angular momentum
(pure relative $S$-state) give larger values. 
In our S-state model for the nucleon, $\Gamma_1^p= 0.278$~\cite{Nucleon2}.

Since a nucleon wave function ($\Psi_N$) dominated by the S-state 
\cite{Nucleon,DIS} overestimates the quark 
contributions to the proton spin, 
we now consider a wave function that 
include also $P$ and $D$-states~\cite{Nucleon2} 
\ba
\Psi_N = n_S \Psi_S + n_P \Psi_P + n_D \Psi_D,
\ea 
where $n_S,n_P$ and $n_D$ are the coefficients 
of the states ($n_S^2 + n_P^2 + n_D^2=1$).
All the components of the wave function
are represented in terms of an off-shell 
quark and two on-shell quarks (quark pair).
We can integrate in the internal degrees of freedom 
of the quark pair and represent the wave function 
in terms of the a quark and a diquark structure
dependent on the nucleon ($P$)
and the diquark ($k$) momenta~\cite{Nucleon2}.
Since in the DIS limit the quarks are pointlike 
the adjustable part of the model 
is restricted to the radial wave functions 
of the states $S,P$ and $D$.
To increase the flexibility of the model 
we also consider different distributions 
(radial wave functions  $\psi_q^S$) 
for the quarks $u$ and $d$.
This asymmetry is supported by the data \cite{DIS,Nucleon2}.

From the calculation of the hadronic tensor,
in which we integrate on the 
quark and diquark on-shell momenta, 
we derive the expressions for the 
DIS structure functions.
In particular the expression for the 
unpolarized structure function 
associated with the $S$-state can be written as
\ba
f_q^S (x) = \frac{M^2}{16 \pi^2} \int_\xi^{+ \infty} 
d \chi  |\psi_q^S(\chi) |^2,
\hspace{.5cm}
\frac{d f_q^S}{dx} = - 
\frac{x(2-x)}{(1-x)^2}  \frac{M^2}{16 \pi^2}  |\psi_q^S(\chi) |^2,
\label{eqFq}
\ea
%where $M$ is the nucleon mass, 
where $\xi= \frac{x^2}{1-x}$ is a function 
of the Bjorken variable $x$, 
and $\chi$ is a covariant variable 
of the nucleon and diquark momenta.
Similar expressions can be written for the $P$ and $D$ components.

Equations (\ref{eqFq}) can be used to conclude 
that the radial wave functions ($L=S,P,D$) can be represented 
in the form
\ba
\psi_q^L (\chi) \propto \frac{\alpha + \beta}{\chi^{n_0} (\beta+ \chi)^{n_1-n_0}},
\label{eqPsiScal}
\ea
where $\alpha$ is a constant,  
$\beta$ is a dimensionless parameter and
$n_0,n_1$ are indices that can be related to 
the values $a_q,b_q$ from 
the parametrizations 
$xf_q(x) \propto x^{a_q} (1-x)^{b_q}$.

To confirm if the CSQM is consistent with 
the DIS regime,  we try to adjust
the parameters of our model to the DIS phenomenology.
Since the experimental data is in some cases 
obtained for very small $Q^2$
(while in the DIS limit $Q^2$ is very large)
we choose to fit our model to the well known 
parametrizations of the data:
Martin, Roberts, Stirling and Thorn (2002) -- {\bf MRST(02)}
(unpolarized structure functions) \cite{Martin02} 
and 
Leader, Siderov and Stamenov (2010)-- {\bf LSS(10)}
(polarized structure functions) \cite{Leader10}. 
We consider the parametrizations for the scale $Q^2= 1 $ GeV$^2$.
%
%The fit is devided in three steps:
We divide the fitting process into 3 steps:
\begin{itemize}
\item
first we estimate the parameters 
of the radial wave functions 
$\psi_q^L$ by a fit to the unpolarized data, $f_u$ and $f_d$,
assuming that all components $S,P,D$ 
have the same shape [see Eq.~(\ref{eqPsiScal})],
\item
based on the first estimate of the radial wave functions
we calculate the  mixture coefficients $n_P$ and $n_D$ 
by making a  fit to the first moment of 
the function $g_1^q$:
$\Gamma_1^u= 0.333 \pm 0.039$, and 
$\Gamma_1^d= -0.335 \pm 0.080$  \cite{DIS,Martin02},
\item
finally the parameters of the radial 
wave functions: $\alpha,\beta$ are adjusted independently 
to the polarized data for 
$\Delta u$ and $\Delta d$.
\end{itemize}

\begin{figure}
\centerline{\mbox{
\includegraphics[height=6cm]{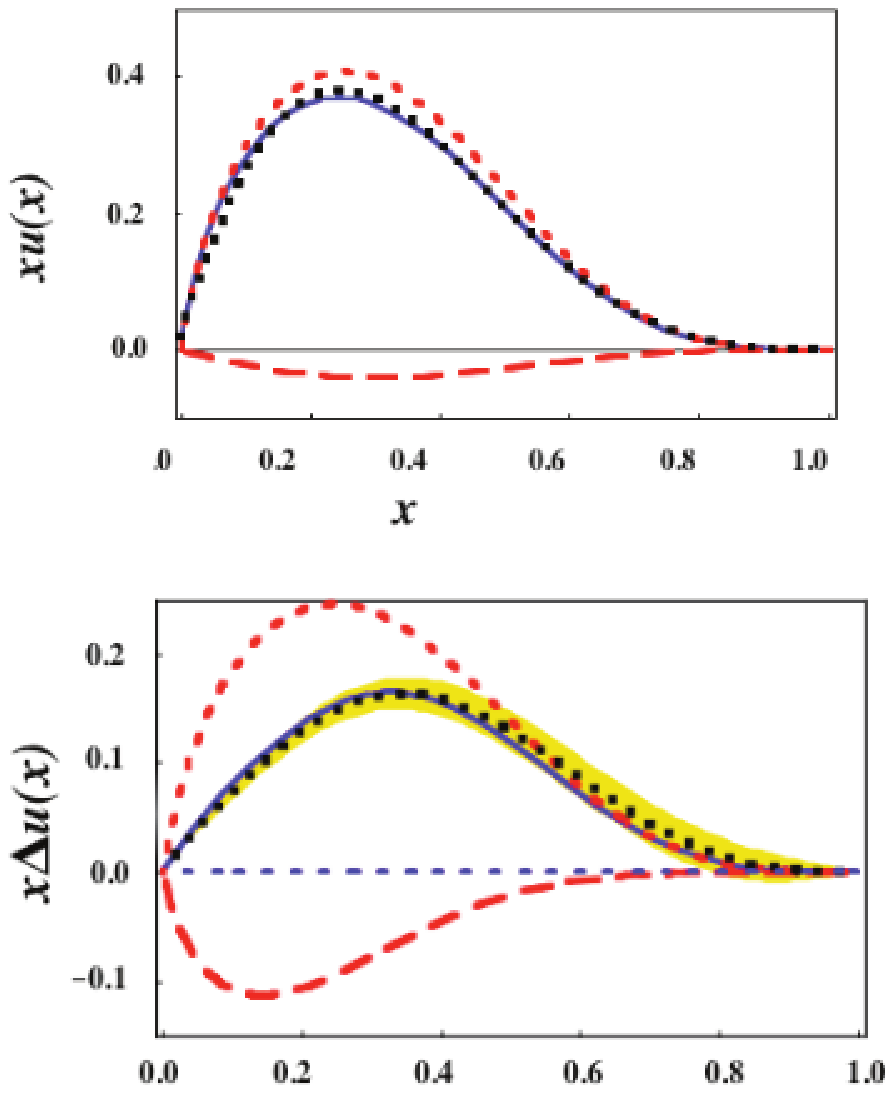} %\hspace{-.4cm} 
\includegraphics[height=6cm]{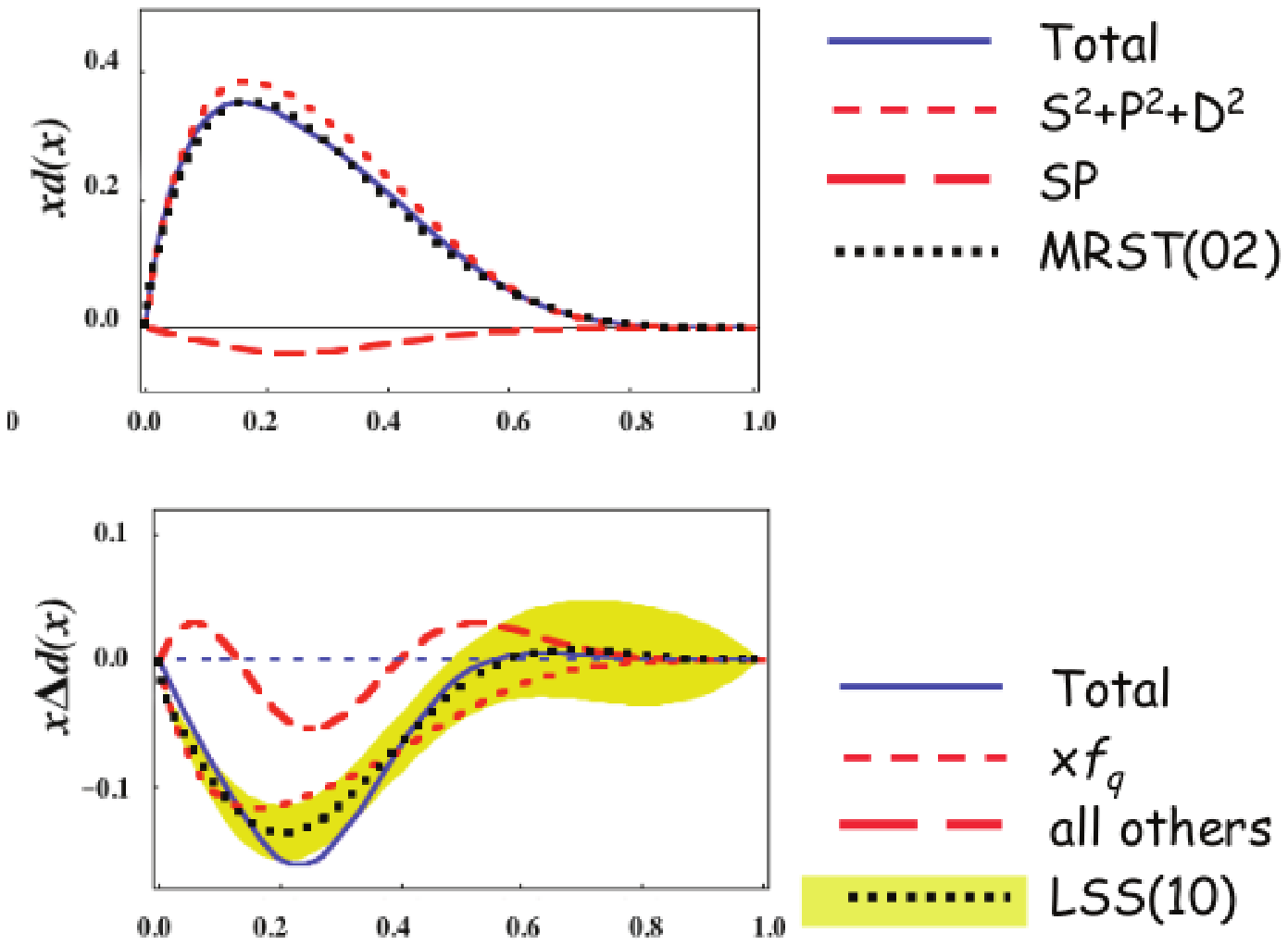} }}
\caption{Results for the unpolarized $q(x)$ 
and polarized $\Delta q(x)$ structure functions (Total)
compared with theparametrizations  MRST(02) and LSS(10) 
\cite{Martin02,Leader10}. 
The $P$- and $D$- state mixtures are respectively 
1\% and 35\%~\cite{DIS}. }
\label{figG1}
\end{figure}

The results of the fit for the functions 
$q$ and $\Delta q$ are presented in the Fig.~\ref{figG1},
and are compared with the parametrizations MRST(02) and LSS(10).

\begin{figure}
\centerline{\mbox{
\includegraphics[width=4.8in]{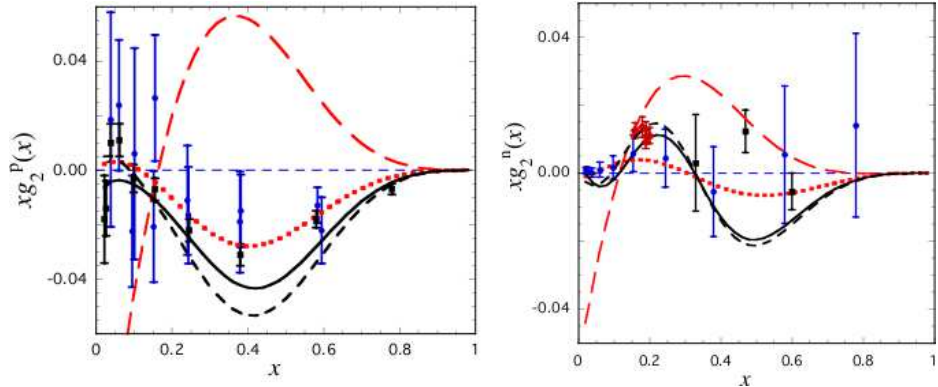} }}
\caption{
Predictions of the function $g_2(x)$
for the proton and neutron (solid line)~\cite{DIS}.}
\label{figG2}
\end{figure}

Once all the parameters are fixed
by the $q$ and $\Delta q$ data, 
we use the model to predict 
the function $g_2(x)$ for the 
proton  and the neutron.
The results are presented in Fig.~\ref{figG2}
by the solid line.

%Write conclusions.

From the previous study 
we conclude that CSQM
be used in the nucleon DIS regime, 
in addition to the electromagnetic excitations of the baryons.
The results presented here are derived under the assumption 
that the valence quarks are the relevant  
degrees of freedom in DIS 
and that the gluon and meson cloud (sea quarks) 
effects can be neglected in a first approximation.

In our study the nucleon has contributions of several 
angular momentum states ($L=S,P,D$)
and the DIS data are used to probe the shape 
of the components of the nucleon wave function. 

The results of our best model 
are consistent with the experimental data
obtained  for the unpolarized $f_q(x)$ and 
polarized $g_1^q(x)$ structure functions,
which are also compatible with a zero contribution 
of the gluons for the proton spin ($J_g=0$).

Finally we present predictions for the 
spin dependent structure function $g_2(x)$
of the nucleon.
The predictions are consistent
with the available data  (see Fig.~\ref{figG2})
and can be tested in future by more accurate data.

Since the gluon degrees of freedom 
are not included explicitly, 
although some effects are effectively considered 
in the structure of the radial wave functions,
we cannot make direct predictions for very large $Q^2$.
We can however use the QCD evolution equations 
(DGLAP) \cite{DGLAP} 
to extrapolate the results to very large $Q^2$,
dominated by the gluon effects,
using the results of our model for  
the valence quark structure at $Q^2=1$ GeV$^2$.

%\newpage


%\vspace{.05cm}

%Acknowledgements:
The author's research was supported
by the Brazilian Ministry of Science, Technology and
Innovation (MCTI-Brazil).

%\input{biblo2}

%%%%%%%%%%%%%%%%%%%%%%%%%%%%%%%%%%%%%%%%%%%%%%%%%%%%%%%%%%%%%%%%%%%%%%
%%%References
%%%%%%%%%%%%%%%%%%%%%%%%%%%%%%%%%%%%%%%%%%%%%%%%%%%%%%%%%%%%%%%%%%%%%%

\end{document}